\documentclass[twocolumn,showpacs,preprintnumbers,amsmath,amssymb]{revtex4}
\usepackage{graphicx}% Include figure files
\usepackage{dcolumn}% Align table columns on decimal point
\usepackage{bm}% bold math

\begin{document}

\preprint{}
\title{Steering far-field spin-dependent splitting of light by inhomogeneous anisotropic media}% Force line breaks with \\
\author{Xiaohui Ling}
%\altaffiliation[ ]{}%Lines break automatically or can be forced with \\
\author{Xinxing Zhou}
\author{Hailu Luo}\email{hailuluo@hnu.edu.cn}
\author{Shuangchun Wen}\email{scwen@hnu.edu.cn}

%%\email{Second.Author@institution.edu}
\affiliation{Key Laboratory for Micro-/Nano-optoelectronic Devices
of Ministry of Education, College of Information Science and
Engineering, Hunan University, Changsha 410082, China}
\date{\today}% It is always \today, today,

\begin{abstract}
An inhomogeneous anisotropic medium with specific structure geometry
can apply the tunable spin-dependent geometrical phase to the light
passing through the medium, and thus can be used to steer the
spin-dependent splitting (SDS) of light. In this paper, we exemplify
this inference by the $q$ plate, an inhomogeneous anisotropic
medium. It is demonstrated that when a linearly polarized light beam
normally passes through a $q$ plate, $k$-space SDS first occurs, and
then the real-space SDS in the far-field focal plane of a converging
lens is distinguishable. Interestingly, the SDS, described by the
normalized Stokes parameter $S_3$, shows a multi-lobe and rotatable
splitting pattern with rotational symmetry. Further, by tailoring
the structure geometry of the $q$ plate and/or the incident
polarization angle of light, the lobe number and the rotation angle
both are tunable. Our result suggests that the $q$ plate can serve
as a potential device for manipulating the photon spin states and
enable applications such as in nano-optics and quantum information.

\end{abstract}

\pacs{42.25.-p, 42.60.Jf, 42.79.-e}% PACS, the Physics and Astronomy
\keywords{spin-dependent splitting, inhomogeneous anisotropic
medium, Stokes parameters $S_3$}

%Use showkeys class option if keyword
\maketitle

\section{Introduction}\label{SecI}
In recent years, some fundamental effects in optics have attracted
much attention, such as the spin Hall effect of
light~\cite{Onoda2004,Bliokh2006}, optical Coriolis
effect~\cite{Bliokh2008}, and optical Magnus
effect~\cite{Dooghin1992}. They manifest themselves as the splitting
of photon spin states, that is, when a linearly polarized light beam
passes through a refractive index gradient (e.g., interfaces of
different media) or an inhomogeneous medium, its left and right
circular polarization components separate from each other. This
spin-dependent splitting (SDS) effect is directly attributed to
different geometrical phases that the two spin components
respectively experienced, corresponding to the spin-orbital
interaction~\cite{Hosten2008,Bliokh2008B}.

The light beam can acquire a spin-dependent geometrical phase upon
the reflection or refraction of a refractive index gradient created
by the interface of different media. When a linearly polarized
paraxial light beam impinges obliquely upon this interface, the SDS
in real space (coordinate space) generates, that is, the two spin
components separate from each other and reside on both sides of the
incident
plane~\cite{Hosten2008,Aiello2008,Luo2009,Qin2010,Hermosa2011,Luo2011A,Zhou2012A}.
This effect is known as the spin Hall effect of light, which has
recently also been extensively studied in other physical
systems~\cite{Kavokin2005,Leyder2007,Maragkou2011,Menard2010,Gosselin2007,Dartora2011},
in addition to optics. Actually, different refractive index
gradients can give a light beam different geometrical phases,
thereby resulting in specified switchable and enhanced SDS
effects~\cite{Luo2011A,Luo2011B}. Therefore, the geometrical phase
can serve as an alternative tool for steering photon spin states. In
general, real-space SDS is often accompanied with the $k$-space
(momentum space) SDS, associated with spin-dependent angular
shift~\cite{Luo2010,Zhou2012B}. Under normal incidence, both
real-space and $k$-space splittings vanish due to the degeneracy of
the geometrical phases of the two spin components~\cite{Hosten2008}.

While for some inhomogeneous anisotropic media, such as surface
plasmonic nanostructures, plasmonic chains, and subwavelength
gratings~\cite{Bliokh2008,Niv2008,Gorodetski2008,Shitrit2011}, even
as under normal incidence, the light beam can also acquire
spin-dependent geometrical phases, and generates the SDS in the
$k$-space. The real-space splitting can be induced after light
propagating to the far field. Some recent researches have also shown
that $q$ plate, a uniaxial birefringent waveplate with space-variant
optical axis orientations, can give a beam $\pm2q$ topological
charge and spin-dependent geometrical phases ($q$ is an integer or a
semi-integer)~\cite{Marrucci2006,Marrucci2011}. Furthermore, tunable
$q$ plates made by liquid crystals have been demonstrated to be
conveniently realized, with arbitrary topological charge and
geometrical phase~\cite{Slussarenko2011,Karimi2009,Piccirillo2010}.
Therefore, from this point of view, it would be interesting to
explore the SDS effect in the $q$ plates as well as its potential
ability to manipulate the photon spin states.

In this work, we theoretically show that the $q$ plate can be
employed to steer the SDS of light in the far field. This is due to
the fact that the $q$ plate can apply a spin-dependent geometrical
phase to a light beam that passes through it. And the geometrical
phase is tunable by varying the $q$ plate geometry or incident
linear polarization angle. When a linearly polarized light beam
normally passes through the $q$ plate, $k$-space SDS first occurs.
After converging by a lens, the induced real-space SDS in the
far-field focal plane is distinguishable. The normalized Stokes
parameter $S_3$ is employed to reveal the separation of spin
photons, because $S_3$ represents the circular polarization degree
of light~\cite{Yariv2007}. We find that the spatial distribution of
$S_3$ exhibits a multi-lobe and rotatable splitting pattern with
rotational symmetry. Further, by tailoring the structure geometry of
the $q$ plate and/or incident polarization angle of light, the lobe
number, and the rotation angle both are tunable. Note that as the
light beam impinges normally into the $q$ plate, the birefringence
does not induce the separation of ordinary and extraordinary light,
which will not contribute to the SDS effect.

\begin{figure}
\includegraphics[height=7cm]{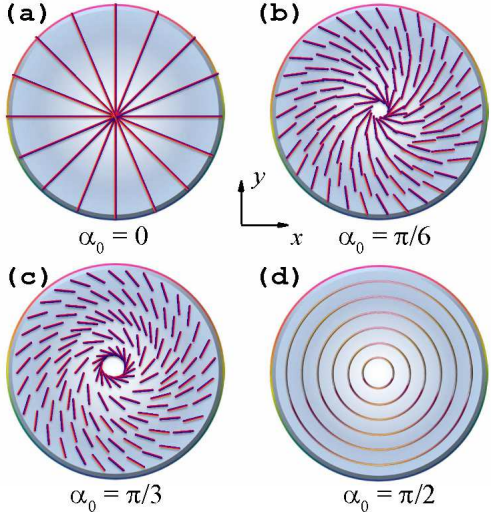}
% Here is how to import EPS art
\caption{\label{Fig1} (Color online) Examples of the $q$ plate
geometries for $q$=1. The tangent to the lines shown indicates the
local optical axis orientation (fast axis). (a)-(d) represent the
geometries for $\alpha_0$=0, $\pi/6$, $\pi/3$, and $\pi/2$,
respectively.}
\end{figure}

\section{Model and theory}\label{SecII}
The $q$ plate is essentially a uniaxial birefringent waveplate with
locally varying optical axis (fast or slow) orientations in the
transverse $xy$ plane, having a homogeneous phase retardation
$\Phi=2\pi(n_e-n_o)d/\lambda$ with $n_e$, $n_o$, $d$, and $\lambda$
the refractive indices of slow and fast waves, material thickness,
and operation wavelength, respectively. It can be currently realized
using liquid crystals, sub-wavelength gratings, or
polymers~\cite{Marrucci2006,Niv2008,Nersisyan2009}. The fast axis
orientations, as specified by a space-variant angle $\alpha(x,y)$ it
forms with the $x$-axis, are described by the following equation:
\begin{equation}
\alpha(x,y)=q\arctan\left(\frac{y}{x}\right)+\alpha_0,
\end{equation}
where $q$ is an integer or semi-integer; and $\alpha_0$ indicates
the angle of local optical axis direction forming with the local
radial direction. Figure~\ref{Fig1} shows four geometries of $q=1$
for different values of $\alpha_0$.

The Jones matrix, describing a uniaxial crystal with a fast axis in
the $x$ direction, can be represented as~\cite{Yariv2007}
\begin{eqnarray}
J=\left[\begin{array}{ccc}
 t_x\exp(i\Phi/2) & 0 \\
 0 & t_y\exp(-i\Phi/2)\end{array}\right],
\end{eqnarray}
where $t_x$ ($t_y$) is the transmission coefficient in the $x$($y$)
direction.

\begin{figure}[b]
\includegraphics[width=8.5cm]{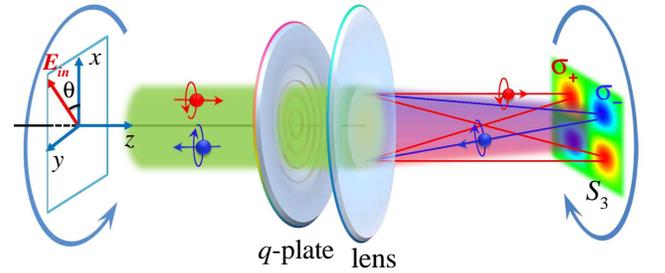}
% Here is how to import EPS art
\caption{\label{Fig2} (Color online) Schematic illustrating the SDS
produced by a $q$ plate. A linearly polarized light beam normally
passes through the $q$ plate, and is focused by a lens. $\sigma_+$
and $\sigma_-$ represent the left and right circular polarization
components, respectively. The two curves with arrows indicate that
the splitting patterns will rotate when rotating the incident
polarization direction ($\theta$). The additional devices for
measuring the Stokes parameter $S_3$ are not shown.}
\end{figure}

\begin{figure*}
\includegraphics[width=15cm]{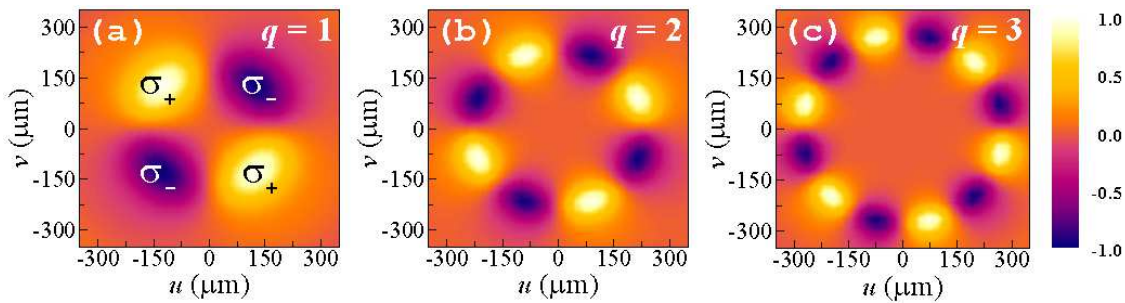}
% Here is how to import EPS art
\caption{\label{Fig3} (Color online) Normalized Stokes parameter
$S_3$ in the focal plane under the irradiation of a linearly
$x$-polarized light for (a) $q$=1, (b) $q$=2, and (c) $q$=3,
respectively. Here, $\alpha_0=0$. In the calculations, we set
$\Phi=\pi/2$, $\lambda$=1 $\mu$m, $w_0$=1 mm, and $f_0$=0.5 m.}
\end{figure*}

As the $q$ plate has space-variant optical axis orientations, a
position-dependent Jones matrix $T(x,y)$ can be employed to fully
characterize the light propagation through the $q$
plate~\cite{Marrucci2006,Yariv2007}:
\begin{eqnarray}
&T(x,y)&=R(-\alpha) J R(\alpha)\nonumber\\
&&=\frac{t_xe^{i\Phi/2}+t_ye^{-i\Phi/2}}{2}\left(\begin{array}{ccc}
 1 & 0\\
 0 & 1\end{array}\right)\nonumber\\
 &&+\frac{t_xe^{i\Phi/2}-t_ye^{-i\Phi/2}}{2}\left(\begin{array}{ccc}
 \cos2\alpha & \sin2\alpha\\
 \sin2\alpha & -\cos2\alpha\end{array}\right),\label{jones}
\end{eqnarray}
where
\begin{eqnarray}
R(\alpha)=\left(\begin{array}{ccc}
\cos\alpha & \sin\alpha\\
 -\sin\alpha & \cos\alpha\end{array}\right).\nonumber
\end{eqnarray}
An input linearly polarized light, with its electric field described
by a Jones vector, is given as
\begin{eqnarray}
E_{in}(x,y)=\left(\begin{array}{ccc}
 \cos\theta\\
 \sin\theta\end{array}\right)E_0(x,y).
\end{eqnarray}
Here, $\theta$ is the linear polarization angle of the polarization
vector forming with the $x$-axis; see Fig.~\ref{Fig2}. Let
$E_0(x,y)$ be a collimated Gaussian beam:
$E_0(x,y)=\exp[-(x^2+y^2)/w_0^2]$ with $w_0$ the beam waist. For
simplicity, we neglect the absorption and loss, and let $t_x$ and
$t_y$ both be equal to 1. Then, the output electric field,
$E_{out}(x,y)=T(x,y)E_{in}(x,y)$, can be calculated as
\begin{eqnarray}
E_{out}(x,y)=\frac{1}{2}e^{-i\theta}\left[\cos\frac{\Phi}{2}+i\sin\frac{\Phi}{2}e^{-i\varphi}\right]\left(\begin{array}{ccc}
 1\\
 i\end{array}\right)E_0\nonumber\\
 +\frac{1}{2}e^{i\theta}\left[\cos\frac{\Phi}{2}+i\sin\frac{\Phi}{2}e^{i\varphi}\right]\left(\begin{array}{ccc}
 1\\
 -i\end{array}\right)E_0, \label{output}
\end{eqnarray}
where
\begin{eqnarray}
\varphi=2\alpha(x,y)-2\theta=2[q\arctan(y/x)+\alpha_0-\theta]\label{phase}
\end{eqnarray}
is an additional space-variant geometrical
phase~\cite{Bomzon2002,Hasman2005}, depending on the local optical
axis orientation of the $q$ plate. $E_{out}(x,y)$ consists of two
parts, corresponding to a coherent superposition of the left and
light circular polarization components, each of which is
respectively composed of two terms: One carries a space-variant
geometrical phase ($\pm\varphi$) and $\pm2q$ topological charge, and
the other does not. This originates from the partial conversion of
the spin-to-orbital angular momentum, with the conversion efficiency
determined by $\Phi$~\cite{Slussarenko2011}. Actually, as the
incident beam is linearly polarized, the partial left-handed photons
transform into right-handed photons and acquire an additional
angular momentum ($+2q\hbar$) to keep the total angular momentum
conserved, and vice
versa~\cite{Marrucci2006,Marrucci2011,Slussarenko2011}. Generally,
the spin-orbital interaction is the origin of such fundamental
effects as the spin Hall effect of light, optical Coriolis effect
and optical Magnus effect, which manifest as the SDS phenomenon.
Thus here, under normal incidence of a linearly polarized light,
$k$-space SDS will first occur, and then real-space SDS upon
propagation. Note that $\varphi$ involves three free parameters,
$q$, $\alpha_0$, and $\theta$, which serve as effective ways for
manipulating the SDS of light.

The momentum shift of the SDS then can be calculated
as~\cite{Bliokh2008,Niv2008,Gorodetski2008,Shitrit2011}
\begin{eqnarray}
&\Delta k&=-\sigma_\pm\nabla\varphi=\Delta k_x+\Delta k_y\nonumber\\
&&=2\sigma_\pm
q\left(\frac{y}{x^2+y^2}\hat{e}_x+\frac{-x}{x^2+y^2}\hat{e}_y\right),
\end{eqnarray}
where $\sigma_+=+1$ and $\sigma_-=-1$, representing the left and
right circular polarization components, respectively, and
$\hat{e}_x$ ($\hat{e}_y$) is the unit vector in the
$x$($y$)-direction. $\Delta k_x$ and $\Delta k_y$ are
position-dependent, with the shift directions respectively
determined by their signs. As the geometrical phase of the $q$ plate
is space-variant both in the $x$ and $y$ directions, the $k$-space
SDS and shift in both directions occur simultaneously. Unlike here,
a recent work reported a constant momentum shift in plasmonic chains
where the SDS effect occurred in one direction~\cite{Shitrit2011}.

\section{Tunable spin-dependent splitting in the far field}
We then consider the far-field SDS effect. A lens with focal length
$f_0$ is used here to converge the output field of the $q$ plate,
and the field in the focal plane can be viewed as a far field, as
schematically shown in Fig.~\ref{Fig2}. The real-space shift induced
by the SDS in the focal plane can be written as $\Delta S=\lambda
f_0\Delta k/2\pi$~\cite{Gorodetski2005}. It is also
position-dependent, which means each point in the $q$ plate
contributes a different momentum shift; thereby, in the far field,
the final splitting effect is determined by their coherent
superpositions.

The electric field distribution in the focal plane can be derived
from the Fraunhofer-diffraction integral formula~\cite{Goodman2005}:
\begin{eqnarray}
E_{out}^{u(v)}(u,v)&=&\frac{\exp(ik_0f_0)\exp\left[\displaystyle
\frac{i k_0}{2f_0}(u^2+v^2)\right]}{i\lambda
f_0}\int\!\!\!\int\limits_{\!\!\!\!\!\!S_{\alpha\beta}}
dxdy\nonumber\\
&&\times E_{out}^{x(y)}(x,y)\exp\left[-\frac{i
k_0}{f_0}(xu+yv)\right], \label{integral}
\end{eqnarray}
where $u$ and $v$ are axes parallel to the $x$ and $y$ axes,
respectively, $S_{\alpha\beta}$ is the space range of the $q$ plate,
and $E_{out}^{x(y)}(x,y)$ represents the $x$($y$)-polarized
component of $E_{out}(x,y)$. As this integral is too complicated to
solve analytically, we will evaluate it numerically.

To show the circular polarization degree of the resulting electric
field and reveal the separation of spin photons, the Stokes
parameter $S_3$ is employed. Describing the SDS by $S_3$ is believed
to be the photonic version of a Stern-Gerlach experiment in the
absence of a magnetic field~\cite{Gorodetski2008,Kang2012}. The
$S_3$, normalized to the total intensity in the focal plane, can be
calculated as~\cite{Yariv2007}
\begin{eqnarray}
S_3=\frac{2|E_{out}^u(u,v)||E_{out}^v(u,v)|\sin(\psi_v-\psi_u)}{|E_{out}^u(u,v)|^2+|E_{out}^v(u,v)|^2},
\end{eqnarray}
where $\psi_{u(v)}$ is the phase of $E_{out}^{u(v)}(u,v)$ and the
denominator represents the total intensity in the focal plane.

\begin{figure}[b]
\includegraphics[width=8.5cm]{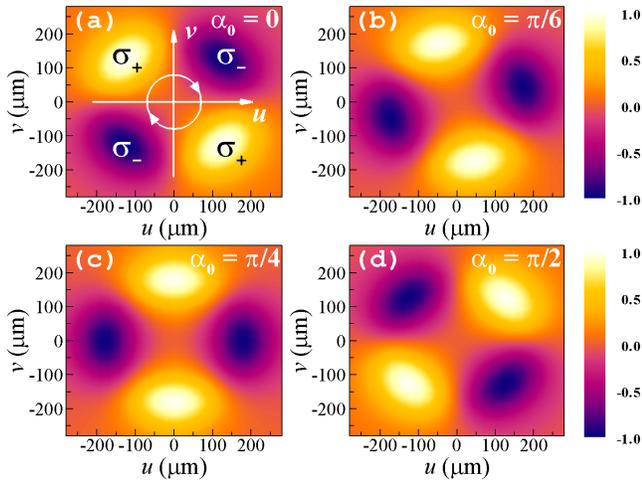}
% Here is how to import EPS art
\caption{\label{Fig4} (Color online) Normalized Stokes parameter
$S_3$ in the focal plane under the irradiation of a linearly
$x$-polarized light for (a) $\alpha_0=0$, (b) $\alpha_0=\pi/6$, (c)
$\alpha_0=\pi/4$, and (d) $\alpha_0=\pi/2$, respectively. Other
parameters are the same as in Fig.~\ref{Fig3}. The white arrows show
the clockwise rotation of $S_3$ with the increase of $\alpha_0$.}
\end{figure}

We now can demonstrate the SDS in the far field produced by the $q$
plates. As stated above, the far-field SDS of left and right
circular polarization components results from their spin-dependent
geometrical phase experienced in the $q$ plate. Since the
geometrical phase is space-variant, the $S_3$ parameter is
inhomogeneous in the far-field focal plane. Note that $\varphi$
involves three tunable parameters: $q$, $\alpha_0$, and $\theta$, we
will then explore the influence of these parameters on the SDS. The
first two parameters are associated with changing the $q$ plate
geometry, and the third one is just related to changing the incident
linear polarization angle which may be more convenient to adjust.

We first discuss the influence of the $q$ value on the SDS.
Figure~\ref{Fig3} shows the spatial distribution of the calculated
$S_3$ parameter for $q=1, 2, 3$ ($\alpha_0=0$) under the linearly
$x$-polarized incidence ($\theta=0$). It is well known that $S_3>0$
corresponds to left-handed polarization helicity and $S_3<0$
corresponds to right-handed polarization helicity. Specifically,
$S_3=+1$ or $-1$ corresponds to $\sigma_+$ or
$\sigma_-$~\cite{Yariv2007}. One can notice that, for $q=1, 2, 3$,
the spatial distributions of $S_3$ show 4, 8, and 12 independent
lobes, respectively, with alternative $\sigma_+$ and $\sigma_-$
components. Moreover, the $S_3$ exhibits twofold rotational ($C_2$)
symmetry for $q=1$, fourfold rotational ($C_4$) symmetry for $q=2$,
and sixfold rotational ($C_6$) symmetry for $q=3$, respectively.
These characteristics are determined by the rotational symmetric
geometries of the $q$ plate and the geometrical nature of the
space-variant geometrical phase [Eq.~(\ref{phase})]. Interestingly,
a similar four-lobe spin splitting effect has also been observed in
semiconductor microcavities for exciton
polaritons~\cite{Kavokin2005,Leyder2007,Maragkou2011} and
subwavelength metallic apertures for photons~\cite{Kang2012}, which
corresponds to the case of $q=1$.

\begin{figure}
\includegraphics[width=8.5cm]{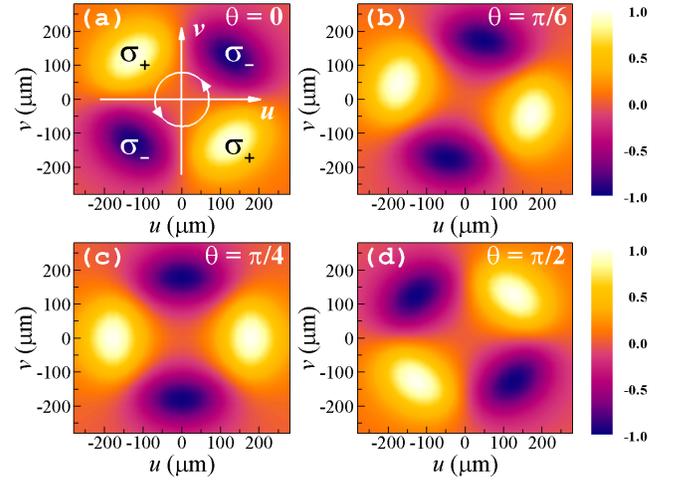}
% Here is how to import EPS art
\caption{\label{Fig5} (Color online) Normalized Stokes parameter
$S_3$ in the focal plane under the irradiation of a linear
polarization light with polarization angle (a) $\theta=0$, (b)
$\theta=\pi/6$, (c) $\theta=\pi/4$, and (d) $\theta=\pi/2$,
respectively. Here, $\alpha_0=0$. Other parameters are the same as
in Fig.~\ref{Fig3}. The white arrows show the counterclockwise
rotation of $S_3$ with the increase of $\theta$.}
\end{figure}

Here, we can give a qualitative explanation for this interesting SDS
phenomenon. Note that the output electric field of the $q$ plate has
two parts with opposite spin handedness and topological charge
($\pm2q$), that is, two helical circular polarization beams [see
Eq.~(\ref{output})]. The wavefront of the helical beam with $\pm2q$
topological charge is composed of $2|q|$ intertwined helical
surfaces, with their handedness determined by the sign of
$2q$~\cite{Marrucci2006}. After interfering and propagating to the
far-field focal plane, these helical surfaces form $4|q|$
independent lobes with alternative $\sigma_+$ and $\sigma_-$
components, which can be described by the $S_3$ parameter.

Nonvanishing $\alpha_0$ and $\theta$ will cause the rotation of the
spatial distribution of $S_3$, since they both can give an initial
value to the geometrical phase $\varphi$. We take $q=1$ for an
example to show the influence of $\alpha_0$ and $\theta$ on the SDS.
As $\alpha_0$ and $\theta$ play similar roles in the geometrical
phase with just opposite signs, their influences will cancel out
each other. When $\theta=0$, increasing $\alpha_0$ will result in
the clockwise rotation of $S_3$ for the $\alpha_0$ radian, as shown
in Fig.~\ref{Fig4}. Accordingly, increasing $\theta$ under the
condition of $\alpha_0=0$ will make $S_3$ rotate counterclockwise
for the $\theta$ radian (see Fig.~\ref{Fig5}). Obviously, $\alpha_0$
and $\theta$ produce the same rotation angles but opposite rotation
directions. These results confirm the predications.

In the above calculations, we have assumed that the $q$ plate served
as quarter waveplates ($\Phi=\pi/2$) with space-variant optical axis
orientations. It is worth noting that, actually, the SDS does not
occur for all cases of $\Phi$. When $\Phi=2m\pi$ ($m$ is an
integer), the term carrying the geometrical phase in
Eq.~(\ref{output}) vanishes, and the remaining part represents a
field with the same polarization as the incident light, that is, no
SDS occurs. When the $q$ plate serves as a half waveplate
$\Phi=(2m-1)\pi$, $E_{out}(x,y)=i[\cos(2\alpha-\theta),
\sin(2\alpha-\theta)]^TE_0$, which represents a linearly polarized
light beam with an axisymmetric polarization
distribution~\cite{Stalder1996}. Thus, for these cases, the SDS
vanishes.

\section{Conclusions}
We have demonstrated a tunable SDS effect in the far field produced
by the inhomogeneous anisotropic media with specified geometries,
called the $q$ plate, under the normal incidence of a linearly
polarized light. This effect originates from the spin-dependent
geometrical phases of the two spin components experienced in the $q$
plate. We have also shown that the SDS, described by the $S_3$
parameter, exhibits a multi-lobe and rotatable splitting pattern,
with the lobe number and rotation angle tunable by the geometrical
phase (associated with the $q$ plate geometry and the incident
linear polarization angle). As the $q$ plate with arbitrary
geometrical phase could be achieved by the current fabrication
technology using liquid crystals, sub-wavelength gratings, or
polymers, we believe that it will serve as a potential device for
manipulating the photon spin states and enables applications, such
as in nano-optics and quantum information.

\begin{acknowledgements}
This work was supported by the National Natural Science Foundation
(Grants No. 61025024 and No. 11074068) and Hunan Provincial Natural
Science Foundation (Grant No. 12JJ7005) of China.
\end{acknowledgements}

\end{document}